# Charging/Discharging Mechanism in $Mg_3Bi_2$ Anode for Mg-Ion Batteries; The Role of the Spin-Orbit Coupling

M. Hussein N. Assadi,[1,*] Christopher J. Kirkham,[2] Ikutaro Hamada,[3] Dorian A. H. Hanaor[4]

[1]*School of Materials Science and Engineering, The University of New South Wales, Sydney, NSW 2052, Australia*
[2]*Center for Materials Research Information Integration (cMI2), Research and Services Division of Materials Data and Integrated System (MaDIS), NIMS, 1-1 Namiki, Tsukuba, Ibaraki 305-0044, Japan*
[3]*Department of Precision Engineering, Graduate School of Engineering, Osaka University, 2-1 Yamada-oka, Suita, Osaka 565-0871, Japan*
[4]*Fachgebiet Keramische Werkstoffe, Technische Universität Berlin, 10623 Berlin, Germany*
[*]h.assadi.2008@ieee.org

Using density functional calculations, we examine insertion/extraction of Mg ions in $Mg_3Bi_2$, an interesting Mg-ion battery anode. We found that a (110) facet is the most stable termination. Vacating a $Mg^{2+}$ ion from the octahedral site is more favourable for both surface and bulk regions of the material. However, the diffusion barriers among the tetrahedral sites are ~ 3 times smaller than those among octahedral sites. Consequently, during the magnesiation/demagnesiation process, Mg ions first vacate the octahedral sites and then diffuse through the tetrahedral sites. The spin-orbit interaction lowers Mg's vacancy formation energy but has a minor effect on diffusion barriers.

The goal of developing new electrochemical storage devices with ever-higher performance in terms of energy storage density, raw material availability and safety has motivated the pursuit of the replacement of Li-ion based systems, which are currently the main type of secondary cell batteries [1,2]. Alternatives such as Na, K, and Mg are currently being considered [3-8]. Among these, magnesium ion batteries (MIBs) [9-13] are particularly advantageous because Mg is a divalent ion offering a relatively high specific capacity of 3833 mA h cm$^{-3}$, or equivalently 2205 mA h g$^{-1}$ [12]. Additionally, Mg has a higher melting temperature than Li (648.8 vs. 180.5 °C), along with a more stable reaction interface, making it a safer and more feasible alternative, relative to Li anodes. Mg is further one of the most abundant materials on earth, offering the potential for scalable battery production. Lastly, Mg does not form dendrites on the anode surfaces, a problem that affects Li-ion systems [14,15]. One of the most significant obstacles to overcome in the long-term use of Mg metal anodes relates to passivation, in which MgO layers or solid electrolyte interphases are formed during operation causing large overpotential as well as low Coulombic efficiency [16].

The overall performance of a magnesium-ion cell dramatically depends on the collective properties of the anode, cathode and electrolyte [17], all of which have design efforts underway. Evidence suggests that through the careful design of electrolytes and anodes, the passivation layer formation can be prevented. In recent studies, the Mg intermetallic compound $Mg_3Bi_2$ has shown promise as an anode material, with a high volumetric capacity, excellent cycling performance and most importantly an electrochemically reversible magnesiation and demagnesiation processes in which no passivation layer is formed [18,19]. However, similar intermetallics, such as Sb or Sb/Bi mixes are not found to be as promising [20], illustrating the need for a deeper understanding into underlying processes. So far, the atomic mechanisms behind Mg vacancy ($V_{Mg}$) formation in such systems and its diffusion processes are not comprehensively understood [21].

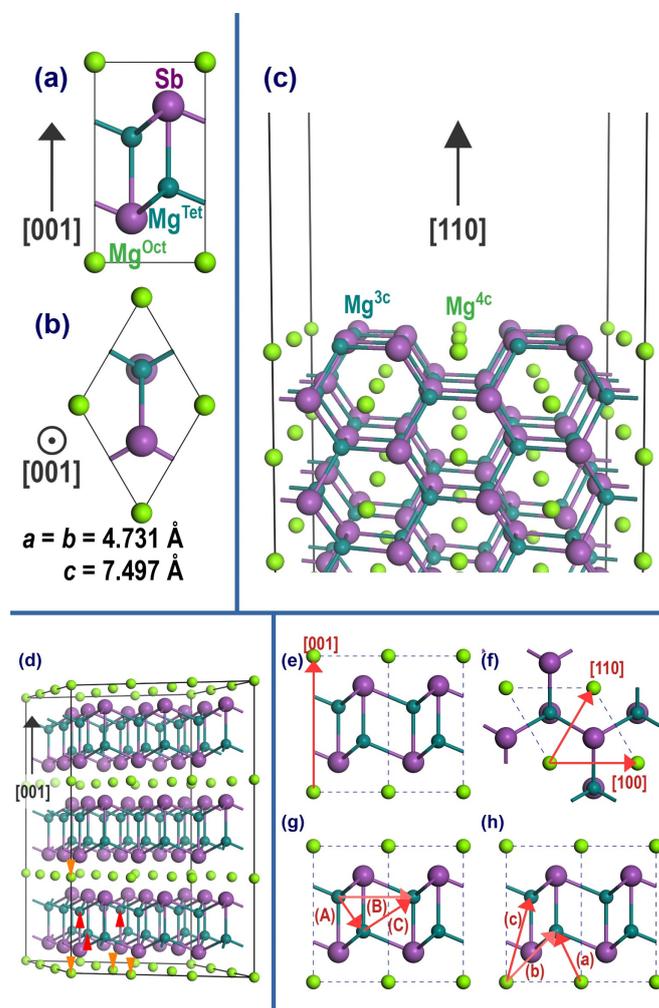

**Fig. 1.** (a)–(b) The side and top views of the bulk unit cell of $Mg_3Bi_2$ with octahedral and tetrahedral Mg sites labelled and colour-coded. (c) The outermost layer of the $Mg_3Bi_2$ cleaved at the [110] direction. (d) The $3a \times 3a \times 3c$ supercell used for calculating the diffusion barriers. The marked Mg ions indicate those vacated during diffusion calculations. The diffusion paths for Mg ions in bulk $Mg_3Bi_2$, as indicated by red arrows; (e)–(f) $Mg^{Oct}$–$Mg^{Oct}$ diffusion, (g) $Mg^{Tet}$–$Mg^{Tet}$ diffusion, (h) $Mg^{Oct}$–$Mg^{Tet}$ diffusion.

The bulk electronic and structural properties of $Mg_3Bi_2$ are well-studied experimentally [22,23] and theoretically [24-27]. However, there is not much information in the literature related to its surface properties, which are essential for understanding the (de)magnesiation process and anode/electrolyte interfaces. In this work, we re-examine the structural and electronic properties of bulk $Mg_3Bi_2$, before considering the energetics of several likely surface terminations. We report the most stable surface orientation and investigate the early stages of demagnesiation at that surface. We present properties relevant to battery operation, such as vacancy formation energies. These results are validated against recent experimental findings.

To study the surface $Mg_3Bi_2$, we performed density functional theory (DFT) calculations with the Vienna *ab initio* Simulation Package (VASP) [28]. The core electrons were described by the projector augmented wave (PAW) method [29]. Calculations were conducted both with and without spin-orbit coupling (SOC) to gain quantitative insight into the effect of Bi mass on the system's electronic structure. The Perdew-Burke-Ernzerhof (PBE) implementation was used for the exchange-correlation functional [30,31]. Further tests were performed with the corrected PBEsol functional [32]. The Bi $6s^2 6p^3$ and Mg $3s^2$ electrons were treated as valence, with the effect of including semi-core Mg $2p^6$ states examined and found to be inconsequential. A 520 eV energy cut-off was used for those calculations without semi-core 2p electrons and 850 eV with semi-core 2p electrons as valence. An $18 \times 18 \times 11$ k-point mesh, or equivalent, was used throughout, keeping the k-point spacing to maximum $\sim 0.012$ Å$^{-1}$.

We only considered the most prevalent hexagonal polymorph of $Mg_3Bi_2$, shown in Fig. 1a–b, as the synthesis of cubic phase material is experimentally challenging and thus of limited interest towards battery applications [23]. The (001), (100) and (110) surfaces were studied using symmetric constructions with up to 51, 31 and 25 atomic layers, respectively, and a 37 Å vacuum slab. The most stable facet is shown in Fig. 1c. Surface vacancies were considered for a 13-layer, $2u \times 2v$ (110) surface. Bulk vacancy formation simulations were considered in up to a $6a \times 6a \times 6c$ supercell. Mg ion diffusion was studied using the Climbing Image Nudged Elastic Band (CINEB) method [33] in a $3a \times 3a \times 3c$ cell, shown in Fig. 1d. For diffusion paths longer than $\sim 4$ Å, 3 intermediate images were considered, while only one image was considered for shorter paths. All atoms were allowed to relax until force components on all atoms were smaller than 0.02 eV Å$^{-1}$. Formation enthalpies ($\Delta H$) were calculated as:

$$\Delta H = \frac{E_{Mg_3Bi_2} - 3 E_{Mg}^{Bulk} - 2 E_{Bi}^{Bulk}}{5} \quad (1)$$

where $E_{Mg_3Bi_2}$ is the total energy of bulk $Mg_3Bi_2$, and $E_{Mg}^{Bulk}$ and $E_{Bi}^{Bulk}$ are the total energies per atom of metallic Mg and Bi phases. Our calculated lattice constants and electronic structure are in good agreement with previously reported measurements; expressly, our lattice constants agree with the experiment within 1% and with only a 0.5% discrepancy for PBEsol [22-27]. PBE+SOC lattice constants ($a = 4.731$ Å, $c = 7.497$ Å) were used in the following, with checks against PBEsol results ($a = 4.665$ Å, $c = 7.359$ Å). Finally, vacancy formation energies ($E^f$) were calculated according to:

$$E^f = E_{Def} + E_{Mg} - E_{Perf} \quad (2)$$

$E_{Def}$ and $E_{Perf}$ are the DFT total energies of the defective and perfect supercells, either in bulk or surface.

Our calculations with SOC yielded $\Delta H = -0.174$ eV for $Mg_3Bi_2$. When SOC is not taken into consideration, $\Delta H$ is calculated to be relatively lower (more stable) as $-0.208$ eV. The inclusion of SOC reduces the calculated stability by $\sim 20\%$, indicating the role of SOC in softening the Mg-Bi bond strength. SOC constitutes a second-order energy term that usually contributes to the total energy [34]. However, if the SOC contribution to the constituent elements in their separate phases is larger than its contribution to the compound, SOC decreases the formation enthalpy [35]. Here, normalised per atom, SOC comprises 6.8% of metallic Bi's total energy, while it comprises only 1.68% of $Mg_3Bi_2$'s total energy. The inclusion of semi-core 2p electrons had a negligible effect ($< 0.5\%$). The inclusion of the Mg semi-core 2p electrons does not significantly influence the structural, energetic, or electronic properties of interest for batteries. Our reported $\Delta H$ of $Mg_3Bi_2$ is in reasonable agreement with experimental measurements [36], but differs significantly from the earlier work by Zhou *et al.*, which reported a formation enthalpy of $\sim -0.738$ eV [25]. This disagreement with earlier calculations is because of looser computational settings used previously, and therefore our $\Delta H$ reporting is more reliable.

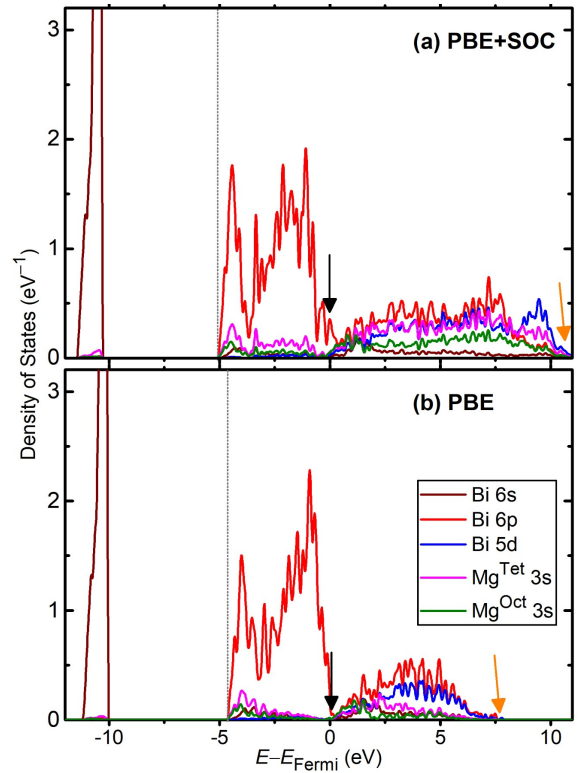

**Fig. 2.** Partial density of states of the pristine $Mg_3Bi_2$. (a) calculated with spin-orbit coupling considered, (b) without the spin-orbit coupling. The origin of the energy is set to the Fermi level ($E_{Fermi}$).

Studying Mg vacancies and their diffusion barriers in $Mg_3Bi_2$ are essential for understanding its behaviour as an anode material. Vacancy formation energies can be related to operating voltages, while diffusion between lattice sites gives a measure of ion mobility. As in Fig. 1a–b, each unit cell of $Mg_3Bi_2$ contains two distinct types of Mg site: octahedral ($Mg^{Oct}$) and tetrahedral ($Mg^{Tet}$) based on their coordination environments. Recently it has been reported that Mg vacancies are relatively mobile [27], but the energetics of the vacancies themselves and their diffusion barriers have not been investigated in detail. According to our calculation with SOC, the formation energy of octahedral Mg vacancy ($V_{Mg}^{Oct}$) and tetrahedral Mg vacancy ($V_{Mg}^{Tet}$) were found to be 0.26 eV and 0.35 eV respectively. These values are ~3 times smaller than those obtained without considering SOC, i.e., $E^f$ = 0.81 eV and $E^f$ = 0.92 eV for vacating an octahedral and a tetrahedral Mg, respectively. The values calculated with SOC compare plausibly with the galvanostatic discharge at 0.25 V, observed in the experiment [18].

The Bader charge analysis for the pristine bulk $Mg_3Bi_2$ primitive cell reveals that Mg ions bear a minor electronic population while Bi ions each bear nearly eight electrons indicating significant charge transfer from Mg to Bi. In supercells with Mg vacancies, the holes created by vacating Mg were partly localised on the neighbouring Bi ions; For $V_{Mg}^{Oct}$, the six-fold coordinating Bi ions each bore 0.136 hole constituting a 0.836 hole, while the rest of the two holes were delocalised on the rest of the Bi ions in the supercell. For $V_{Mg}^{Tet}$, the four coordinating Bi ions each bore 0.167 hole while the rest of the holes were delocalised over the other Bi ions.

The partial density of state (DOS) of $Mg_3Bi_2$ is presented in Fig. 2. Due to relativistic contraction, Bi 6s states constitute a separated narrow band at $-12$ eV $< E <$ $-10$ eV. By comparing the Bi 6s states calculated with SOC (Fig. 2a) and without SOC (Fig. 2b), we see that this situation is not affected much by SOC. This is because s orbitals do not contribute to the intra-atomic SOC. Moreover, the separation of Bi 6s states renders them electrochemically inert [37]. Bi 6p states occupy a wide valence band that starts at $E = -5$ eV and stretches to higher energies beyond the Fermi level. The majority of the Bi 6p states nonetheless gravitate towards the bottom of the valence band below the Fermi level. With SOC, as marked with dotted lines in Fig. 2, Bi 6p band is slightly wider, and the magnitude of the lower energy states is slightly larger. The $Mg^{Oct}$ and $Mg^{Tet}$ 3s states ultimately coincide with Bi 6p states facilitating the charge transfer from Mg to Bi. Our calculation with SOC predicts a non-zero DOS at the Fermi level, marked with black arrows, agreeing with earlier experimental measurements [38]. However, without SOC, the DOS at the Fermi level is negligible, hinting at an extremely narrow bandgap compound. Moreover, with SOC, the conduction band is predicted to be wider, stretching to ~10 eV above the Fermi level, as marked with orange arrows.

To properly characterise $Mg_3Bi_2$ as a battery material, it is essential to identify the most stable surface termination or facet. Here, as shown in Fig. 3, we considered surfaces along three lattice directions, (001), (100) and (110), with five, three and one unique terminations, respectively. To assess the relative stability of each surface we compared their surface energy per unit area ($\gamma$). Since most surfaces were non-stoichiometric, we used:

$$\gamma = \frac{E_{Slab} - n E_{Mg_3Bi_2} - m \mu_{Mg}}{2A}, \quad (3)$$

where $E_{Slab}$ is the energy of the slab, $n$ is the closest number of complete bulk unit cells compared against the slab, $m$ is the number of excess or deficient Mg atoms in the slab, and can be positive or negative accordingly, $\mu_{Mg}$ is the Mg chemical potential, and $A$ is the surface area. For stoichiometric slabs $m = 0$. $\mu_{Mg}$ is constrained between Bi/Mg rich limits such that:

$$\frac{1}{3} E_{Mg_3Bi_2} \leq \mu_{Mg} \leq E_{Mg}. \quad (4)$$

Four different criteria were used to examine the convergence of the slab properties with respect to thickness. The surface relaxation limit (SRL) is when surface structural parameters, such as bond lengths, surface buckling and inter-layer relaxation are converged to within 1 %. The thin film limit (TFL) is when inter-layer relaxation at the centre falls below 0.1 %, such that there is no deviation from the bulk. The surface energy limit (SEL) is when the surface energies are converged to within chemical accuracy, 1 meV Å$^{-2}$. The bulk energy limit (BEL) is when the energy to add two layers converges to within 0.1 % of the energy of two bulk unit cells. The limits for all nine surface terminations are presented in Table 1.

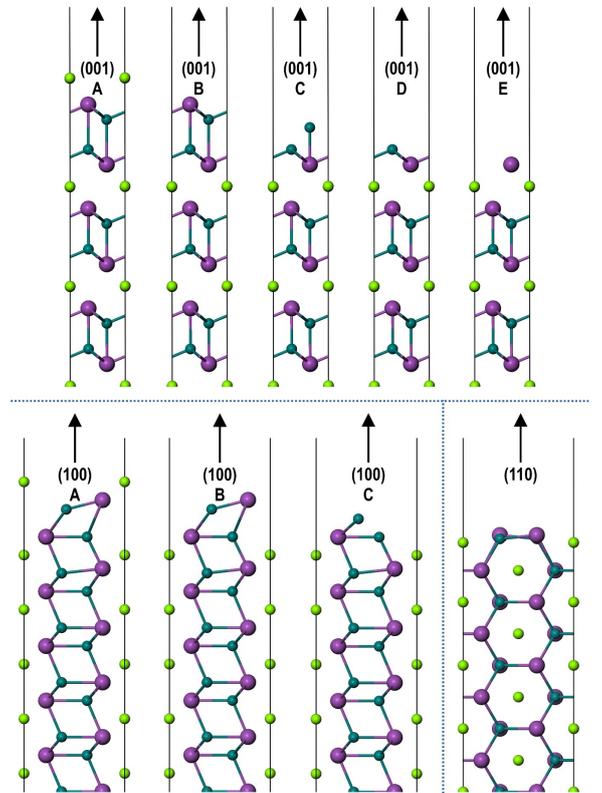

**Fig. 3.** Relaxed surface structures along the (001), (100) and (110) directions, with possible terminations, labelled. Purple, light green and dark green spheres represent Bi, octahedral Mg and tetrahedral Mg ions, respectively.

The significant differences in required thickness between orientations is related to the thickness of their bulk equivalents; five, three and one atomic layers for (001), (100) and (110) respectively. The thicker the bulk equivalent, the thicker the slab must be. In all cases, TFL is the most stringent limit, which except for the (110) surface, is impractical for further calculations. All three of the remaining criteria are satisfied by 31 layers for (001), 23 layers for (100) and 11 layers for (110), with slight reductions depending on the surface termination. For the thicknesses considered here, SRL could not be reached for the (001) E termination, with convergence to $< 2\%$ by 43 layers.

Well converged surface energies for all 9 surfaces are presented in Fig. 4. For the full range of the Mg chemical potential considered here, the (110) surface was found to have the lowest surface energy, with Termination B of the (100) surface coming within 1 meV Å$^{-2}$ at the Bi rich limit. Reexamining the surface energies with the PBEsol functional was found to change the absolute values by $5\sim 8$ meV Å$^{-2}$ but did not affect the relative stabilisation sequence of the surfaces for the range of $\mu_{Mg}$ considered. Our theoretical prediction of the (110) surface stability corroborates earlier experimental observation indicating the (110) surface being the preferred growth direction [39]. Thus, the (110) is the most stable surface and was considered here for further calculations.

**Table 1.** The required number of atomic layers for the convergence of different $Mg_3Bi_2$ surface terminations.

|  | Limit (Atomic Layers) | | | |
| --- | --- | --- | --- | --- |
| Surface | SRL | TFL | SEL | BEL |
| (001)A | 31 | 51 | 21 | 31 |
| (001)B | 19 | 49 | 19 | 29 |
| (001)C | 27 | 47 | 17 | 27 |
| (001)D | 15 | 45 | 15 | 25 |
| (001)E | — | 43 | 23 | 23 |
| (100)A | 19 | 25 | 19 | 13 |
| (100)B | 23 | 29 | 17 | 17 |
| (100)C | 15 | 27 | 15 | 21 |
| (110) | 9 | 13 | 11 | 9 |

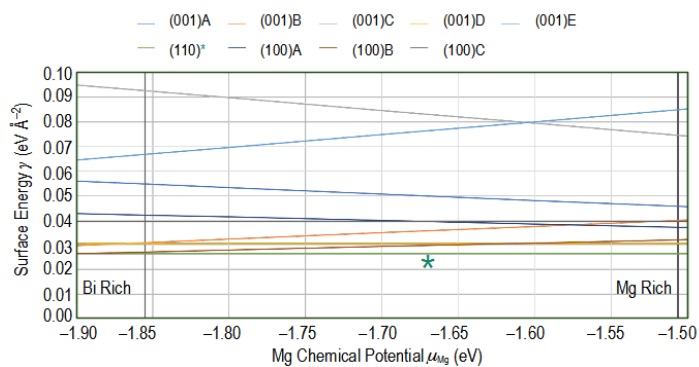

**Fig. 4.** Surface energies for 9 different surface terminations of $Mg_3Bi_2$, between Bi-rich (lower) and Mg-rich (upper) limits. For the entire range of the permissible Mg chemical potential, the (110) surface, as marked with an asterisk, is the most stable.

With the (110) surface established as the most stable termination; we explored the initial stages of magnesium extraction from $Mg_3Bi_2$ anodes by considering Mg vacancy formation on the outermost layer of the (110) surface. There are two distinct Mg varieties at this surface, either as part of a Mg-Bi dimer or in the space between dimers. These are what used to be the tetrahedral and octahedral Mg sites before cleavage, respectively. As in Fig. 1c, we mark these Mg ions with $Mg^{3c}$ and $Mg^{4c}$, respectively. The superscripts indicate their coordination number at the (110) surface. The formation energy of $V_{Mg}^{4c}$ and $V_{Mg}^{3c}$ at the (110) surface were 0.15 eV and 0.18 eV, respectively with SOC considered. As in bulk, these values are substantially smaller than those obtained without considering the SOC (0.32 eV and 0.43 eV for $V_{Mg}^{4c}$ and $V_{Mg}^{3c}$, respectively, at the (110) surface). This trend suggests Mg leaves the $Mg^{4c}$ sites first.

The Bader charge analysis for the (110) surface rendered similar results to the bulk system. The $2u \times 2v$ stochiometric (110) supercell contained 156 Mg and 104 Bi atoms, respectively. Before vacating a Mg, all Mg ions were found to bear nearly zero Bader charge while all Bi atoms found to each, on average, bear 8 $e$, including the Bi atoms at the outermost layers of the surface. The creation of $V_{Mg}^{4c}$ or $V_{Mg}^{3c}$ at the surface resulted in mostly delocalised holes that reduced the Bader charge of the outermost Bi atoms by only $\sim 0.02$ $e$, while the rest of the holes were delocalised over the Bi ions within the deeper layers. Consequently, the Bi atoms at the surface retain an anionic character (similar to their counterparts in the bulk), both in magnesiated and demagnesiated states.

We considered the Mg diffusion between equivalent and inequivalent lattice sites along different lattice directions in the bulk environment to identify the preferred path Mg ions travel along for reaching the surface. The results are shown in Table 2. The paths of diffusion are depicted in Fig. 1e–h. Between the $Mg^{Oct}$ sites (Fig. 1e–f), we found the barrier along the longer [001] was lower at 0.66 eV than the shorter paths of [100] and [110]. This trend did not noticeably change without SOC, even though the values with SOC were slightly smaller. Between $Mg^{Tet}$ sites (Fig. 1g,), we found that diffusion between nearest sites was preferred, regardless of direction, with barriers of only 0.27 eV, which was only 0.01 eV smaller than the barrier calculated without SOC. Between $Mg^{Oct}$ and $Mg^{Tet}$ sites (Fig. 1h), we found that diffusion to the nearest site was still greatly favoured over the next nearest, with barriers of 0.30 and 0.68 eV, respectively. Noticeably, $Mg^{Oct}$ vacancies are 0.11 eV lower in formation energy than $Mg^{Tet}$ vacancies, implying Mg diffusion from $Mg^{Tet}$ to $Mg^{Oct}$ will have 0.11 eV higher barriers. Consequently, Mg vacancies are formed at the octahedral sites, and afterwards, diffuse rapidly from $Mg^{Oct}$ sites to $Mg^{Tet}$ sites, and then back and forth between neighbouring $Mg^{Tet}$ sites through the shortest path, with the same barriers for both processes. This mechanism is similar to the formation and diffusion predicted without SOC, except for the vacancy formation energy and diffusion barriers are smaller. The calculated low Mg vacancy formation energies and diffusion barriers qualitatively agree with the experimentally observed fast Mg kinetics and low activation barrier (~0.19 eV) [40].

Strong spin-orbit coupling, such as that which occurs in $Mg_3Bi_2$, generally reduces the cohesive energy and raises the compound's formation enthalpy ($\Delta H$). Larger bulk formation enthalpy, in turn, reduces the formation energy of defects. This reduction is further magnified in the Mg vacancy, as SOC does not significantly raise the formation energy of light-weight Mg. A similar correlation between low defect formation energy and low bulk formation enthalpy has been previously reported in the literature [41]. The SOC effect on the diffusion barriers, however, was less drastic. The reason is that the diffusion of Mg ions is mainly determined by the electrostatic repulsion from negatively charge Bi ions [42]. Electrostatic effects are typically several times larger in magnitude than SOC. Therefore, the inclusion of SOC in the calculations does not significantly alter the diffusion barriers [43].

To compare to what extent the mass-dependent SOC affects the formation energy of the Mg vacancies, we calculated the formation energy of $V_{Mg}^{Oct}$ and $V_{Mg}^{Tet}$ in $Mg_3Sb_2$. $Mg_3Sb_2$ is isomorphic and isovalent to $Mg_3Bi_2$, but the mass of Sb is ~60% lighter than that of Bi. Without considering the SOC, the formation energy of $V_{Mg}^{Oct}$ and $V_{Mg}^{Tet}$ were found to be 1.48 eV and 1.49 eV, respectively. When the spin-orbit coupling interaction was included in the calculations, the formation energies were reduced to 1.25 eV for $V_{Mg}^{Oct}$ and 1.26 eV for $V_{Mg}^{Tet}$. As expected, the effect of the SOC interaction in $Mg_3Sb_2$ was very modest compared to that in $Mg_3Bi_2$. Consequently, Mg compounds with larger SOC seem more suitable for anode application to facilitate the extraction/insertion of the Mg ions.

In summary, we reported our theoretical investigations into structural and electronic properties of bulk $Mg_3Bi_2$ and explored the properties of several different surface terminations. We demonstrated that a (110) facet containing two Mg ion types—$Mg^{3c}$, and $Mg^{4c}$—is the most stable termination for $Mg_3Bi_2$. On this facet, vacating $Mg^{4c}$, which is the most coordinated Mg ion, is more favourable. Mg vacancy formation within the bulk region is similarly more favourable from octahedral Mg sites than tetrahedral sites. However, the diffusion barriers among the octahedral sites are ~3 times larger than those among tetrahedral sites.

Consequently, during the (de)magnesiation process, Mg ions first vacate $Mg^{4c}$ or $Mg^{Oct}$ sites. The Mg vacancies, in turn, move to and diffuse through the tetrahedral sites. The spin-orbit coupling significantly lowered the Mg vacancy formation energy but had only a minor effect on the diffusion barrier energies.

The authors gratefully acknowledge the funding of this project by computing time provided by the Paderborn Center for Parallel Computing (PC²).

**Table 2.** The diffusion barriers calculated without and with SOC. The paths of diffusion are marked with red arrows in Fig. 1e–h. Distances quoted correspond to the distance between the initial and final location of the Mg ion in the pristine supercell.

| Diffusion Path | PBE+SOC (eV) | PBE (eV) | Distance (Å) |
|---|---|---|---|
| $Mg^{Oct} \rightarrow Mg^{Oct}$ [100] | 0.76 | 0.77 | 4.73 |
| $Mg^{Oct} \rightarrow Mg^{Oct}$ [110] | 0.76 | 0.77 | 4.73 |
| $Mg^{Oct} \rightarrow Mg^{Oct}$ [001] | 0.66 | 0.75 | 7.49 |
| $Mg^{Tet} \rightarrow Mg^{Tet}$ (A) | 0.27 | 0.28 | 3.32 |
| $Mg^{Tet} \rightarrow Mg^{Tet}$ (B) | 0.27 | 0.28 | 3.32 |
| $Mg^{Tet} \rightarrow Mg^{Tet}$ (C) | 0.66 | 0.68 | 4.73 |
| $Mg^{Oct} \rightarrow Mg^{Tet}$ (a) | 0.30 | 0.39 | 3.11 |
| $Mg^{Oct} \rightarrow Mg^{Tet}$ (b) | 0.30 | 0.39 | 3.11 |
| $Mg^{Oct} \rightarrow Mg^{Tet}$ (c) | 0.68 | 0.97 | 3.89 |